%% file: krim.tex
\documentclass[12pt]{iopart}
\usepackage{iopams}
\usepackage{epsfig}

\def\ead#1{\vspace*{5pt}\address{E-mail: \mailto{#1}}}
\def\mailto#1{{\tt #1}}


 \newcommand{\bc}{\begin{center}}
\def\ec{\end{center}}
\newcommand{\Pom}{\mbox{$I\!\!P$}}
\newcommand{\aP}{\mbox{$\alpha_{\Pom}$}}
\begin{document}

\title
{Proton Structure Function Measurements    
   from HERA
}

\author{J\"org Gayler
}

\address{DESY, Notkestrasse 85, 22603 Hamburg, Germany}
\ead{gayler@mail.desy.de}

\begin{abstract}
Measurements of proton structure functions made
 in neutral and charged current interactions at HERA
are discussed,
 covering
four-momentum transfers $Q^2$ from about
0.5 GeV$^2$ to $30\,000$ GeV$^2$.
The results include
the rise of the structure function $F_2$ towards small $x$ and
 electro-weak effects
at high $Q^2$.
 QCD fits made by the H1 and ZEUS collaborations
 provide both, parton densities with
uncertainties, and precise $\alpha_s$ determinations.
\end{abstract}

\input{paper}




\end{document}

%% file: paper.tex
\section{Introduction}
The proton is probably 
 the most studied hadron. Whereas general parameters
 like the mass are measured to an accuracy of about $10^{-7}$, the internal
 properties are known at best at the few percent level.
 The internal structure, as probed in hard interactions,
 is described in terms
 of parton density functions (pdfs).
 These are determined, in particular, in lepton nucleon scattering experiments.
 Such
 measurements\footnote{Presented at XXXII
   International Symposium on Multiparticle Dynamics,
   Alushta, Crimea,  \\
 7. - 13th September 2002} 
     are important for two reasons, they provide an important testing
  ground for QCD, but also because  
   the pdfs 
  are needed to make predictions
     for other reactions, e.g. $\bar{p}p$ collisions.  
 
 In inclusive $e^+p$ ($e^-p$) scattering the proton structure can be probed
 by $\gamma$ or $Z^0$ exchange, i.e.
 by neutral current (NC) interactions ($ep \rightarrow eX$),
 or by $W^+$ ($W^-$) exchange, i.e.
 by charged current (CC) interactions ($ep \rightarrow \nu X$).
 The NC differential cross section can be expressed in terms of
 three structure functions, $\tilde{F_2}$, $\tilde{F_3}$ and $\tilde{F_L}$:  
\begin{equation}
d^2 \sigma^{\pm}_{NC}/dxdQ^2 =
\frac {2\pi \alpha^2}{xQ^4} [ Y_+ \cdot \tilde{F_2}
         \;   \mp \; Y_- \cdot x\tilde{F_3}  -y^2
      \cdot \tilde{F_L}] \equiv \frac {2\pi \alpha^2}{xQ^4}
       \tilde{\sigma}^{\pm}_{NC}\;,
\end{equation}
where  $Y_{\pm} = 1 \pm (1 - y)^2$. Here, $Q^2=-q^2$ with $q$ being
 the four-momentum
 of the exchanged gauge boson,  $x = Q^2/2(P\cdot q)$,
 the momentum fraction of the proton carried by the parton
 participating in the interaction, and
 $y = (P\cdot q)/(P\cdot k)$,
 the
 inelasticity, where $k (P)$ is the
 four-momentum of the incident electron (proton).
 The structure function $\tilde{F_2}$ is the dominant contribution in
 most of the phase space and in leading order (LO) QCD
 can be written in terms of
 the quark densities
 $\sim x \sum_q e_q^2(q(x) + \bar q(x))$.
 The term  $x\tilde{F_3}$ contributes
  significantly\footnote{the $\tilde{F_i}$
 contain also $M_Z$ terms originating from $Z$ exchange.}
 at $Q^2 \gtrsim M_Z^2$
 and is to LO  $\sim  x \sum_q (q(x) -\bar q(x))$, that is,
  it is given by the
 valence quarks.
The longitudinal
 contribution $\tilde{F_L}$ is important in Eq. (1) only at large $y$.
 At small $x$, to order $\alpha_s$,  $\tilde{F_L} \sim \alpha_s g$,
 where $g$ is the gluon density.
 
 Similarly, the CC cross section can be written
\begin{equation}
d^2 \sigma^{\pm}_{CC}/dxdQ^2 =
\frac {G^2_F}{2\pi x} (\frac{M^2_W}{Q^2 + M_W^2})^2
  \cdot    \tilde{\sigma}^{\pm}_{CC}\;\;,
\end{equation}
where $G_F$ is the Fermi coupling constant.
 
In LO
$\;\;\;\tilde{\sigma}^+_{CC} = x[(\bar u(x) + \bar c(x)) + (1 - y)^2(d(x) + s(x))]$\\
 \hspace*{32pt} and
$\;\;\;\tilde{\sigma}^-_{CC} = x[(u(x) + c(x)) + (1 - y)^2(\bar d(x) + \bar s(x))]$.\\
 The $d$-quark density is therefore directly accessible
 in $e^+p \rightarrow \bar \nu_e X$ scattering
 avoiding the nuclear corrections necessary in electron 
 deuteron scattering.
 
\section{Electro-weak Effects}
 
\begin{figure}[ht]
\begin{picture}(200,200)
   \put(-8,-10.){\epsfig{file=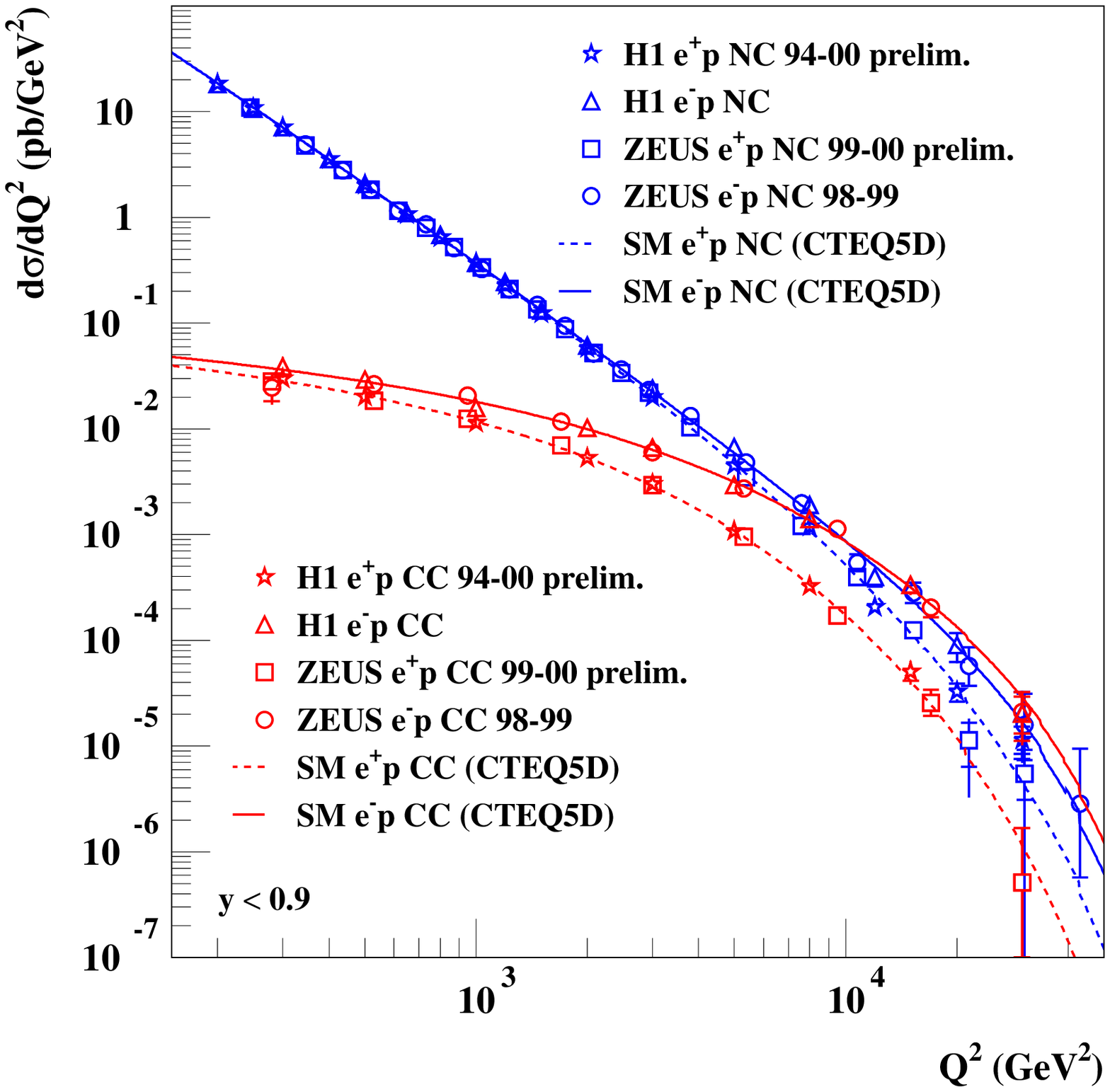,width=230pt}}
  \put(219,30.){\epsfig{file=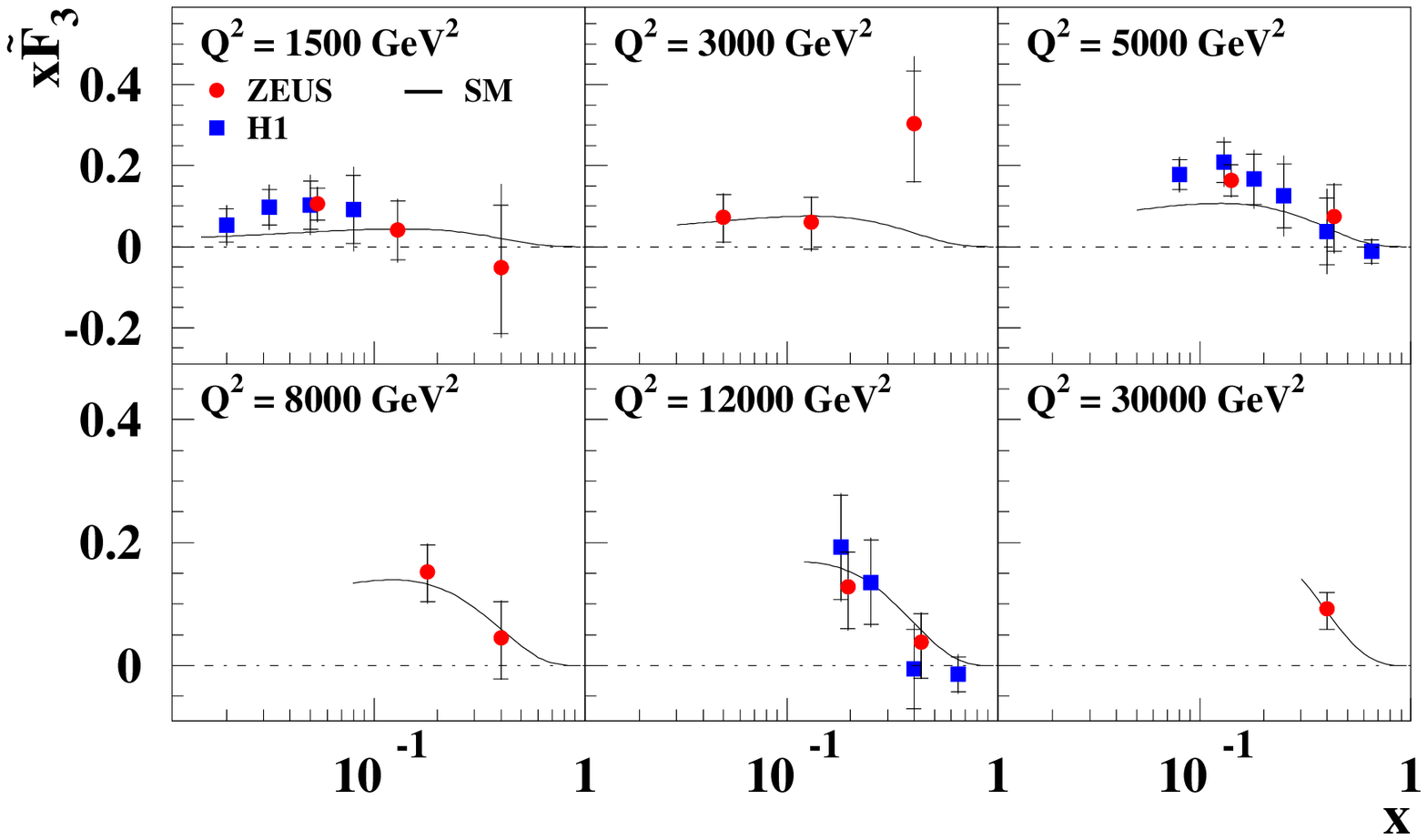,width=220pt}}
 \end{picture}
\caption{
 a) $d\sigma/dQ^2$ for $e^+p$ and $e^-p\;$ NC and CC interactions compared
with the standard model (SM) based on CTEQ5D$\;^1$ pdfs,
b) $x\tilde{F_3}$ vs. $x$ compared with
SM expectation$\;^2$.
\label{fig:dsq}}
\end{figure}

  The $e^+p$ and $e^-p$ data on NC and CC interactions at high $Q^2$
  are summarised in Fig.~\ref{fig:dsq}a.
 In NC, $\rm d \sigma/\rm d Q^2 \sim 1/Q^4$ due to photon exchange.  
At $Q^2 \approx 100$ GeV$^2$
 the cross section is about a
  factor 1000 larger than the CC
  cross section which varies as $\sim 1/(Q^2 + M_W^2)$.
  However, we observe that at $Q^2 \gtrsim M_Z^2, M_W^2,\;\;
  \sigma_{CC} \approx \sigma_{NC}$ illustrating  electro-weak unification
  in deep inelastic scattering (DIS).
 
 A closer look at Fig.~\ref{fig:dsq}a shows that the $e^-p$
  cross sections are above those of $e^+p$. In the CC case,
  this follows from the valence contribution which is
  $\sim u_v(x)$
   for $e^-p$ scattering and $\sim (1-y^2)\cdot d_v(x)$ for  $e^+p$.
In the NC case, this difference is seen in more detail in
 Fig.~\ref{fig:dsq} b)
which shows $x\tilde{F_3}$ which is dominated by the $\gamma Z^0$ interference
 term.
 Taking the electro-weak couplings into account,
 $x\tilde{F_3}^{\gamma Z} \sim 2u_v + d_v$. Future, more precise HERA
 measurements of $x\tilde{F_3}$ will provide an interesting consistency check
 for the valence quark densities based on NC $ep$ scattering only.

\section{Recent QCD Analyses of DIS data}

 In the standard DIS QCD analyses a parameterisation
 of the pdfs at a starting scale $Q_0^2$ is assumed,
 which are evolved to higher $Q^2$ using the
 NLO DGLAP equations~\cite{Furmanski:1980cm}.
 The parameters at $Q_0^2$ are determined
 by a fit of the calculated cross sections or $F_2$ values to the
 data.
 The analyses differ mainly in the amount of data used, the handling
 of systematic errors, the parameterisations at $Q_0^2$, and the treatment
 of heavy quarks. Results of
 such analyses were recently presented by H1 and ZEUS, leading to
 pdfs with associated uncertainties.

\begin{figure}[ht]
\begin{picture}(140,290)
\put(120.,0.)
{\epsfig{file=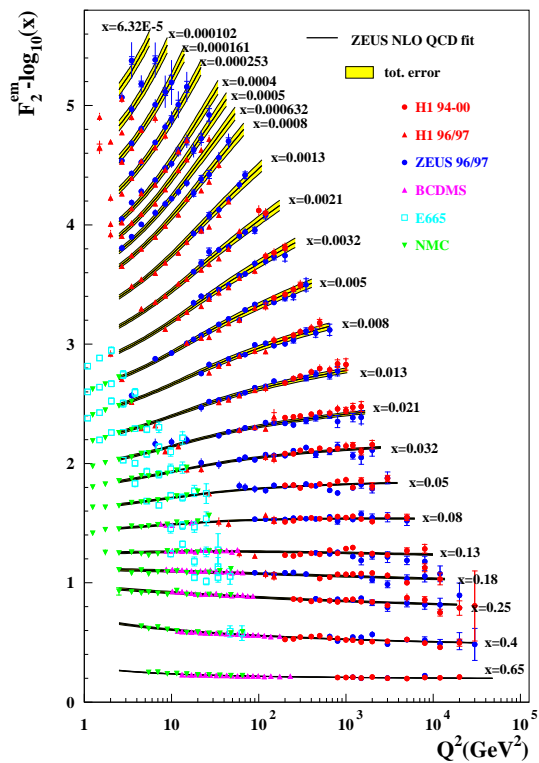,width=200pt}}
 \end{picture}
\caption{
 $F_2^{em}$, i.e, $F_2$ due to $\gamma$ exchange, 
 from HERA and fixed
 target experiments 
  compared with the
the ZEUS NLO fit~\cite{Chekanov:2002pv}.
\label{fig:fitdata}}
\end{figure}
 The H1 2000 QCD fit~\cite{Adloff:2000qk}
 used the H1 $ep$ NC data and BCDMS $\mu p$ data.
The primary purpose was
 a determination
 of the gluon density $g(x)$
 and the strong
coupling constant $\alpha_s$.
 For this reason,
 besides $g(x)$ only two
 functions were parametrised  at  $Q_0^2$,
  one for the
 valence
 and one for the sea quark contribution, with small corrections.

 The preliminary H1 2002 pdf fit~\cite{h1pdf}, which includes in addition
 the H1 CC and the BCDMS $\mu d$ data, determines $g(x)$ and also the
 four up and down combinations
  $U = u+c,\;\; \bar U = \bar u + \bar c,\;\; D = d + s,\;$ and
  $\bar D = \bar d + \bar s\;\;$ from which the valence densities
  $u_v = U - \bar U$ and $d_v = D - \bar D$ are derived.
  Fitting the H1 data alone gives essentially the same pdfs, but with
  increased uncertainties at large $x$.
  In this case the sensitivity to
  $d(x)$ is mainly due to the $e^+p$ CC data.
 
 The recent ZEUS analysis~\cite{Chekanov:2002pv} uses
 ZEUS NC data, $\mu p$ and $\mu d$ data from BCDMS, NMC and E665,
  and CCFR $\nu Fe$ data. Results on $g(x),\; u_v(x)$, $d_v(x),\;$
 the total sea and $\bar d - \bar u\;$ are given.

 The ZEUS and H1 NLO fits describe the data very well
 (Fig.~\ref{fig:fitdata}).

 The fits follow the steep rise of $F_2$ at small $x$ which is driven
 by $g(x)$. The question remains whether the
 DGLAP approach is good enough at small $x$ where
 $\alpha_s \ln 1/x$ terms are neglected.
  The parameterisations for the
 $x$ dependence at $Q_0^2$ are indeed flexible,
  but the $Q^2$ dependence of the data is well described by DGLAP evolution
  without further parameters.

\begin{figure}[ht]
\begin{picture}(190,190)
\put(30.,0.)
{\epsfig{file=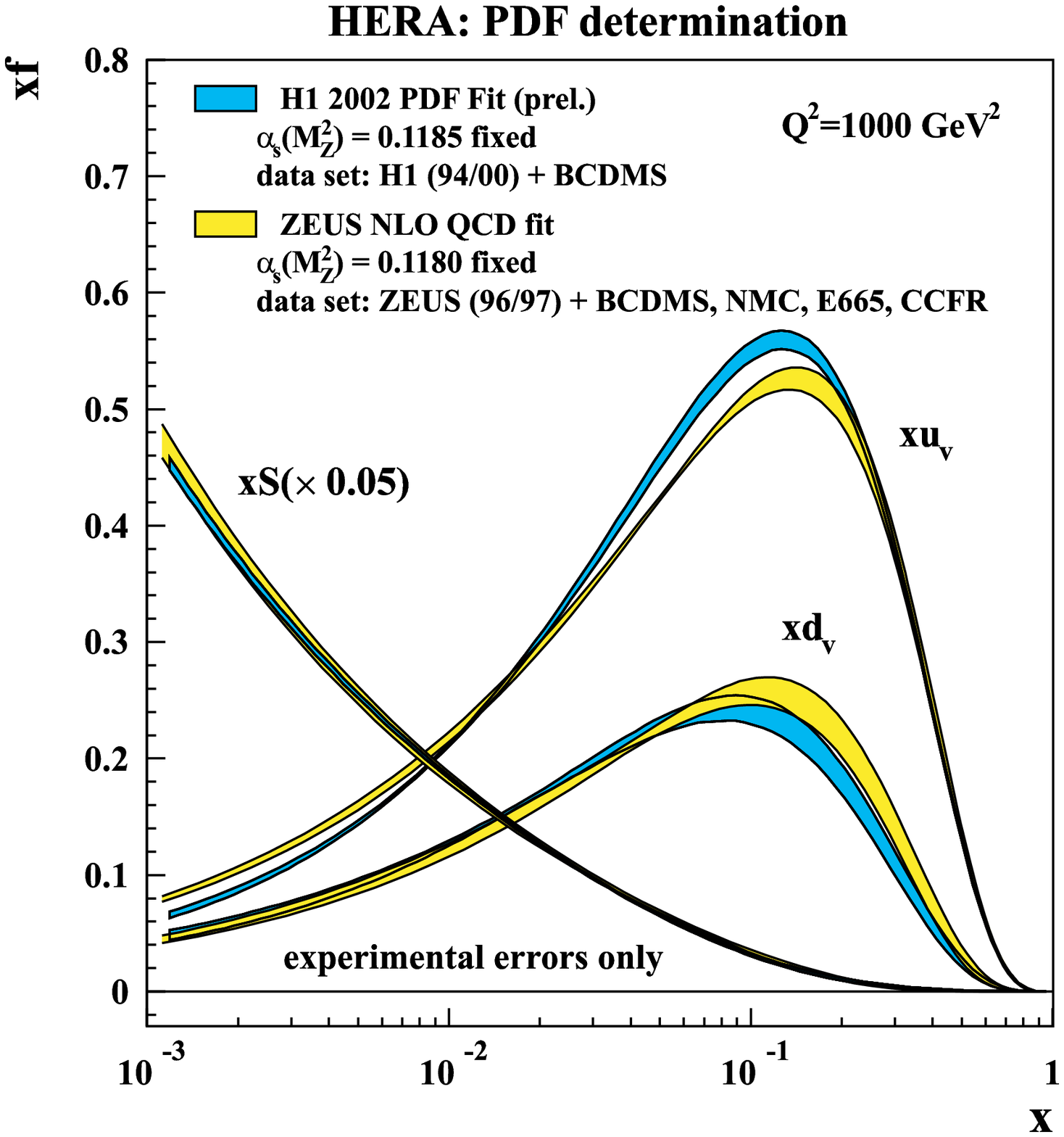,width=176pt}}
\put(233.,-2.)
{\epsfig{file=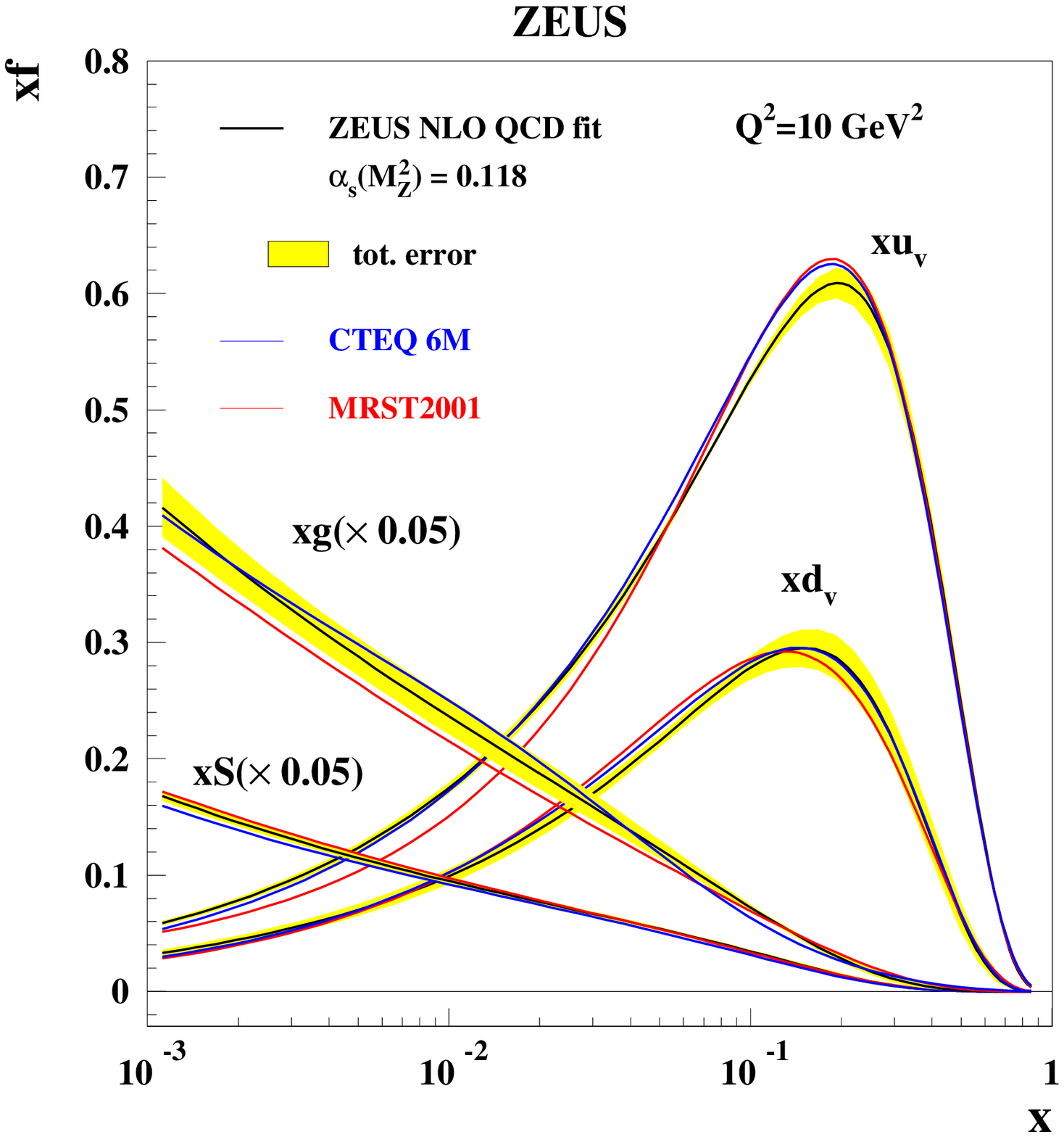,width=178pt}}
 \end{picture}
\caption{
a) Comparison of pdfs of
 the prel. H1 2002 pdf fit~\cite{h1pdf} with
 the ZEUS NLO fit~\cite{Chekanov:2002pv}
b) comparison of the ZEUS fit
 with the global analyses CTEQ6M~\cite{Pumplin:2002vw}
 and MRST2001~\cite{Martin:2001es}.
\label{fig:zh1pdfs}}
\end{figure}

 The resulting pdfs of the fits are compared in Fig.~\ref{fig:zh1pdfs}.

The H1 and ZEUS results are consistent at the 5 to 10\% level
 and also agree
 with the results of
 global analyses~\cite{Martin:2001es,Pumplin:2002vw}.
 This is remarkable
 in view
 of the different methods and the different data sets used.

 The strong rise of the gluon density towards small $x$ leads to
 the prediction of a substantial
$F_L$
 contribution to the cross section
 which is
 consistent with the data~\cite{Adloff:2000qk}.

In the central H1 and ZEUS fits,  $\alpha_s(M_Z^2)$  is kept fixed.
If treated as a free parameter, the results
$\alpha_s(M_Z^2) = 0.1150 \pm 0.0017 {\rm (exp)} \; ^{+0.0009}_{-0.0005}
 {\rm (model)}$  (H1 2000 QCD fit~\cite{Adloff:2000qk}) and
$\alpha_s(M_Z^2) = 0.1166 \pm 0.0008{\rm (uncorr.)} \pm 0.0032 {\rm (corr.)}
 \pm 0.0036{\rm(norm.)} \pm 0.0018{\rm(model)}$
   (ZEUS analysis~\cite{Chekanov:2002pv}) are obtained,
 which are competitive with other $\alpha_s$ determinations.
 However,
 theoretical uncertainties due to missing
 higher orders are estimated to be $\approx \pm 0.005$~\cite{Adloff:2000qk}.
 This uncertainty are expected to be considerably reduced by full
 next to NLO calculations which are expected to be  completed
 soon~\cite{Moch:2002sn}.

\section{The rise of $F_2$ towards low $x$}
 
 The rise of the proton structure function $F_2$ towards small $x$
  has been discussed already in the early days of QCD.
  In the double asymptotic limit
  (large energies, i.e. small $x$, and large photon virtualities $Q^2$)
  the DGLAP evolution equations can be solved~\cite{DeRujula:rf}
   and $F_2$ is
   expected to rise approximately like a power of $x$ towards low $x$.
  Power like behaviour is also expected in the BFKL approach ~\cite{bfkl}.
  However, it was soon realised~\cite{Gribov:1981ac}
  that this rise must eventually be limited to satisfy  unitarity constraints,
  perhaps as a result of gluon fusion in the nucleon.
  Experimentally, the rise towards small $x$ was
 first observed
  in 1993 in the HERA data~\cite{Abt:1993cb}.

\begin{figure}[ht]
\begin{picture}(140,210)
   \put(100.,-4.){\epsfig{file=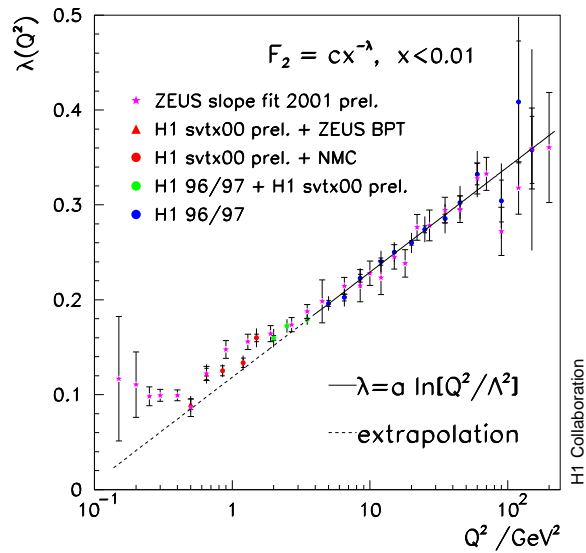,width=220pt}}
 \end{picture}
\caption{
$\lambda(Q^2)$ from fits of the form 
$F_{2} = c(Q^2) \cdot
 x^{-\lambda(Q^{2})}$
 (results from refs. 
 \cite{Adloff:2001rw},\cite{zeusph}).
\label{fig:lammerge}}
\end{figure}

  Now the improved precision of the data
 allows detailed study of the rise through the determination of
  $\lambda \equiv -(\partial \ln F_2/\partial \ln x)_{Q^2}$ as a function of
  $x$ and $Q^2$.
  The derivative $\lambda$ was shown~\cite{Adloff:2001rw}
  to be constant within
  experimental uncertainties at $x < 0.01$ for fixed $Q^{2}$ in the range
  $0.5 \lesssim Q^2 \lesssim 150$ GeV$^2$,
  implying that the data are consistent with the behaviour
  $F_{2} = c(Q^2) \cdot x^{-\lambda(Q^{2})}$.
  Fitting this form to the HERA and fixed target data at
  $x < 0.01$, results in
 $\lambda$ values  (Fig.~\ref{fig:lammerge}) which
 rise logarithmically for
 $Q^2 \gtrsim 3.5$~GeV$^2$,
 that is in the region
 where perturbative QCD fits are thought to be valid.

\vspace*{2pt}
 At small $Q^2$ the structure function $F_2$ can be related
 to the total virtual
 photon absorption cross section by
$
\sigma_{tot}^{\gamma^*p} = 4 \pi \alpha^2 F_2 /Q^2 \; \sim x^{-\lambda}/Q^2\;,
$
 where the total $\gamma^*p$ energy squared is given by $s = Q^2/x$.
 For $Q^2 \rightarrow 0$ we can expect
   $\lambda(Q^2) \rightarrow 0.08$. This corresponds
   to the energy dependence observed in soft hadronic interactions
   $\sigma_{tot} \sim s^{\aP(0)-1}$ with $\aP(0)-1 \approx 0.08$
~\cite{Donnachie:1992ny}
   which is approximately reached at $Q^2 \approx 0.5$~GeV$^2$.

\section{Conclusion}
    New improved  data on inclusive $\;e^{\pm}p$ 
    scattering have become available in recent years. 
 At high $Q^2$, NC and CC interactions are
 consistent with the expectations of electro-weak theory and QCD. 
 
H1 and ZEUS have performed DGLAP based pQCD analyses
which describe their data very well
and
 provide pdfs including uncertainties.
The strong coupling $\alpha_s$
 was determined with good experimental accuracy.
 
At low $x$, no significant deviation from a power behaviour
    $F_{2} \sim x^{-\lambda}$ at fixed $Q^2$
 is visible at
present energies and $Q^{2} \gtrsim 0.85$ GeV$^{2}$.
At  
 $Q^2 \lesssim 1$~GeV$^2$, the rise with energy is similar
 to that observed in soft hadronic interactions.

\section*{Acknowledgments} 
 I am grateful to 
 Vladimir Chekelian, Malcolm Derrick,
 Tim Greenshaw, and Zhiqing Zhang for comments.